*On the shoulders of students*?
The contribution of PhD students to the advancement of knowledge

Vincent Larivière*

École de bibliothéconomie et des sciences de l'information,
Université de Montréal, Montréal, Québec, Canada
and
Observatoire des sciences et des technologies, Centre interuniversitaire de recherche sur la science et la technologie, Université du Québec à Montréal, Montréal, Québec, Canada

* Address for correspondence
Vincent Larivière
École de bibliothéconomie et des sciences de l'information
Université de Montréal
C.P. 6128, succursale Centre-ville
Montréal (Québec)
Canada H3C 3J7

Phone: +1.514.343.5600
Fax: +1.514.343.5753
Email: vincent.lariviere@umontreal.ca

**Abstract**

Using the participation in peer reviewed publications of all doctoral students in Quebec over the 2000-2007 period this paper provides the first large scale analysis of their research effort. It shows that PhD students contribute to about a third of the publication output of the province, with doctoral students in the natural and medical sciences being present in a higher proportion of papers published than their colleagues of the social sciences and humanities. Collaboration is an important component of this socialization: disciplines in which student collaboration is higher are also those in which doctoral students are the most involved in peer-reviewed publications. In terms of scientific impact, papers co-signed by doctorate students obtain significantly lower citation rates than other Quebec papers, except in natural sciences and engineering. Finally, this paper shows that involving doctoral students in publications is positively linked with degree completion and ulterior career in research.

**Introduction**

Graduate students are an important part of the academic workforce. In Canada, more than 32,000 students were enrolled in PhD programs for the year 2003, and more than 3,800 graduated the same year (Canadian Association for Graduate Studies, 2006). Of those students, about 30 % were studying in Quebec. There are almost as many doctoral students as full-time university professors in Canada—40,800 in 2006 (Association of Universities and Colleges of Canada, 2007). In the province of Quebec, about 27,500 distinct students were enrolled in doctoral programs at some point between 2000 and 2007 (GDEU database, compilations performed by the author), while there were 9,306 university professors in the province in 2006 (CREPUQ, 2009).

It is often stated that PhD students are in an intermediary position, in which they need to acquire new skills and knowledge, but also to contribute to the advancement of knowledge in their scientific discipline (Delamont, Atkinson and Parry, 1994). As mentioned by Ziman (1993, cited in Sadlak, 2004, p.8), "…the PhD experience is the psychological transition from a state of being instructed on what is already known to a state of personally discovering things that were not previously known". This is, ultimately, the purpose of doctoral programs: to form new researchers who can contribute to the advancement of knowledge and, during the process, form other new doctoral students. It is during this period of their lives that graduate students get socialized into research; that they acquire the behaviours, attitudes, norms and know-how's of their scientific community.

Socialization to research is very complex and includes a wide spectrum of activities, such as lab work, meeting with advisors, writing of research proposals, staff meetings, etc (Nettles and Millets, 2006). One of these components is the publication activity, which can be defined as the process by which newly created scientific knowledge goes, through peer review, from the lab to the scientific community. Publication can be considered to be a very important component of the socialization and integration of students to research, because it is through that process that knowledge gets validated—or rejected—by the scientific community. Given the increased competition of academia today, the mere writing of a thesis, though necessary, cannot be considered as a sufficient condition to enter today's scientific community. In other words, it is considered that the complete training of doctoral researchers does not end with the thesis, but includes the publication of its results in the scientific community (Kamler, 2008).

Hence, given that a central component of the research habitus is to publish new knowledge—in scientific papers or books—, it is in large part by participating in that publication and diffusion effort that graduate students are socialized into research and integrated into the scientific community, not by the sole writing of a thesis. Though students defend their theories/discoveries/ideas in their theses, which they submit to various committees of peers and, hopefully, future colleagues, this exercise takes place in a 'controlled atmosphere', given that the committee's composition is often decided with the collaboration of their advisors—the latter having an interest in seeing their students graduate. Hence, the thesis submission and defence can be considered more as a rite of passage (Bourdieu, 1982; Goffman, 1974), as opposed to the standard peer review that is experienced when submitting a paper to a journal. When graduate students submit papers to scientific journals, they are in the real scientific world and have no control over the choice of their evaluators. It is this form of peer review that will become the rule if they pursue a career in research after they graduate.

Beyond these general considerations, little is known on the extent of graduate students' contribution of the creation of new knowledge. Though we know that less than half of all graduate students aim for a university career (Fox and Stephan, 2001) and that only a third of Quebec's PhD graduates become integrated into faculty (Conseil Supérieur de l'Éducation, 2003) we have as yet no data on PhD students' participation in the production of new knowledge. In order to shed light on this question, this paper presents the results of the first large-scale survey of PhD students' publication activity, using the whole population of PhD students enrolled in Quebec's universities between 2000 and 2007 (N=27,393). It first provides a measure of the extent to which graduate students participate in the publication process during their studies and, hence, are integrated and socialized into the research activity. In addition to the extent of this socializing practice, this paper also investigates into some of the context in which these papers are produced, as well as its relationship with their subsequent research careers. Finally, this paper assesses the contribution of graduate students to research within Quebec's research system, which includes in this paper all domains of scholarly activity (natural sciences, medicine, social sciences, arts and humanities).

**PhD students' publication activity**

Very few studies have attempted to measure graduate students' participation in the production of knowledge. As Nettles and Millett (2006) put it, "[a]lthough students are believed to acquire preparation for their lives as scholars and researchers while attending graduate school, evidence of their research productivity during their doctoral students days is not abundant. [...] The dissertation [...] is the only research product for which there is comprehensive documentation" (p.105). This section reviews the literature on students' contribution to research, as measured by their contribution to published scientific literature.

From the qualitative analysis of a physics laboratory, Shinn (1988) showed that the research results of junior researchers had a greater cognitive value than those of senior researchers and, because of their greater precision, were often able to end scientific controversies. Also studying experimental physics, Walford (1983) came to the conclusion that students make "a significant contribution to research" (p.253), though these contributions are often of a technical nature. Nevertheless, no study has yet attempted to measure that contribution quantitatively at the macro level. The main reason for that is the technical difficulty associated with the identification of the authors and their status (professor, student, postdoctoral researcher, etc.), which are discussed in detail in the *methods* section).

Because of these technical limits, most studies conducted so far are small scale case studies focusing either on a small sample of students or on specific fields.

Also, given the difficult task of obtaining an assessment that is objective and independent of the actors' own opinions of their contributions—advisors and students often disagree on their respective roles in student research (Berelson, 1960; Campbell, 2003)—most of these studies were made using bibliometrics. For the field of information sciences, Anwar (2004) examined the pre-doctoral (1991-1995) and post-doctoral (1996-2000) publication activity of 54 individuals who graduated from U.S. universities in 1995. His data showed that 24 graduate students out of 54 (44.4%) published one document or more (journal article, conference paper, book or annual review) during the course of their PhD, while a third of all graduates have not published anything over the ten-year period studied.

For the field of adult education, Blunt and Lee (1994) studied, using survey data, the contribution of students to papers published in the *Journal of Adult Education/Adult Education Quarterly* over the 1969–1988 period. Their data shows that 113 students contributed—either as authors or co-authors—to 128 articles published in the journal, accounting for a major share (46%) of all papers published in the journal. Male students accounted for 69% of all students' papers, while females participated in 31% of the papers—though at an increasing rate over the period studied. The authors also analyzed collaboration trends of students: 55% of the papers were written by one student alone, 39% had two authors and 6% had more than two authors. These collaborators were either students or non-students.

In the field of medicine, Cursiefen and Altunbas (1998) examined the research output of medical students of a German medical faculty over the 1993–1995 period. Using Medline, they found that students were the authors of 28% of the entire faculty's paper (316 out of 1128) and the first authors in slightly less than half of the cases. Similarly, Whitley, Oddi and Terrell (1998) studied students' participation in the publication process for the field of nursing, surveying the 633 authors publishing in the journal *Nursing Research* between 1987 and 1991. Their data showed that 31.6% of all authors publishing in that journal for that period were students. Though their survey did not include any information on students' formal collaboration with their supervisors for the paper(s) they published in the journal *Nursing Research*, 84.5% of the students surveyed indicated that they were supervised by Faculty during the process.

In an attempt to provide data for a spectrum of subfields, Lee (2000) analyzed the participation in papers for a sample of PhD students in analytical chemistry, experimental psychology and American literature. Using *Dissertation Abstracts* to build a sample of PhD students, he then compiled publication data for seven cohorts of students graduating between 1965 and 1995. As might be expected, Lee's data presented interesting differences among the three fields: while about 85% of 1995 graduates in analytical chemistry published at least one paper during their studies, this percentage dropped to 50% in experimental psychology and 35% in American literature. Lee also shows that, at least for the fields of analytical chemistry and American literature, there was an increase in student participation in the publication process over the period. In terms of collaboration trends, the author observed a decline in solo authorship in all three fields similar to what is observed at the macro-level.

Apart from the micro-level studies described above, the most second large-scale study on the topic is that of Nettles and Millett (2006), who, in 1996, surveyed more than 9,000 doctoral students in the U.S. who had completed at least one year of study in their PhD program. This major survey analyzed several dimensions of the PhD experience—financing, socialization, satisfaction, etc—among which scientific *productivity* is one component. Their results show that about one graduate student in two had published some of their research—be it as a conference paper, an article, a book chapter or a book—during their studies. More specifically, 66% of engineering students, 57% of humanities students, 52% of science and mathematics students, 47% of social science students and 40% of education students had some research output. This data is not, however, broken down by both discipline and document type; the high figures—especially in the humanities—could be caused by the inclusion of conference papers and posters, which can be considered as more 'entry-level' publications than journal articles or book chapters.

For instance, if only journal articles are used to measure doctoral students' research output, 47% of engineering students and 44% of science students had some scientific production to their name, while only 22% of students in the social sciences, 19% of students in the humanities and 15% of students in education did so. Their data also showed that there were some variations between fields and students' cultural background (the authors use race as indicator) and that research productivity was positively linked with degree completion and time spent in the program, i.e., students with a publication record did not spend more time in the program. Moreover, the authors show that, in all fields but education and social sciences, men were, on average, publishing more papers than women.

Gemme and Gingras (2008) conducted a similar survey, albeit on a smaller scale (104 respondents), on the socialization into research of graduate students (at the master's and PhD level) in Quebec. Though their dataset is not large enough to compile meaningful statistics on publication practices of students by field or level of study, their data nonetheless show that slightly more than 80% of students had contributed to at least one publication since the beginning of their graduate program. Indeed, 55% of students had contributed to at least one conference paper, 43% to at least one research report, 41% to at least one article and 39% to at least one poster. Very few students contributed to books (2%) or book chapters (5%).

In sum, these mostly small scale studies only provide a partial view of student participation in the publication process. Though they all follow a valid method—survey or bibliometrics—they are generally too narrow in scope (one or a couple of subfields). In fact, even if we were to combine these studies, which we cannot given the different methods used in each of them, we would not obtain a general portrait, since subfield studies only represent a small fraction of the scientific world. Taken individually, these studies nonetheless tend to indicate that students do contribute to scientific publications and that, in some fields, their contribution is substantial.

**Factors affecting students' publication activity**

The main source of information on the factors affecting doctoral students' participation to peer reviewed papers found in the literature is the very large scale survey of doctoral students performed by Nettles and Millets (2006), which provides unique results on the effect of funding and supervisors on students' participation. The authors present the different type of funding offered to students during their doctorate, broken down into three categories: fellowships, research and teaching assistantships. As one could expect, students of the different disciplines are not equals in terms of access this research funding. While 69% of students in humanities received fellowships, this percentage was of 61% in social sciences, 59% in sciences and mathematics, 50% in the engineering and 46% in education. Their study does not provide any indication on the amount received, as smaller fellowships might explain the high percentage of students funded in the humanities.

In terms of research assistantships, the tendency is quite different: 82 % and 69%, respectively, of students in engineering and in science and mathematics worked as faculty members' assistants, while this percentage was only 49%, 33% and 28% in the social sciences, humanities and education. Finally, in all disciplines except education, a majority of students received teaching assistantships. This funding was found to have a strong effect on the PhD students' participation in peer-reviewed papers. In education, science/mathematics and social sciences, students receiving fellowships published more papers than those who did not receive any. Similarly, in all disciplines but the humanities, students who were research assistants published more papers. Teaching assistantship was positively linked with research productivity only in the humanities. Along the same lines, Buchmueller, Dominitz and Hansen (1999) have also positively linked students' research assistantships with research productivity, using a sample of doctoral students in economics. Working with productive faculty members also increased students' research output.

Nettles and Millets (2006) analysis of faculty-student interaction revealed that these interactions were more frequent for students who expect their first job to be faculty or postdoctoral position. Similarly, those who declared having mentors also had more faculty-students' interactions. Mentoring was also found to have a strong effect on research productivity. This positive effect—students with mentors published more papers—was observed in all disciplines.

Paglis, Green and Bauer (2006) also analyzed the relationship between mentoring and research productivity for 161 'hard' sciences doctoral students. Using submitted papers instead of papers actually published, they found that 'collaborative' mentoring—defined as a mentoring relationship in which the students are asked by their mentors to be the co-authors of papers, conference papers, books, book-chapters or research grant proposal—was significantly associated with research productivity. This means that co-authorship with advisors/mentors had an influence on students' publication activities.

In a survey of doctoral students in Quebec, Gemme and Gingras (2008) analyzed the importance of supervisors in the selection of their research problem. The authors have found that supervisors of more than 92% of respondents had been involved in the definition of the students' research projects. This percentage was below average in pure sciences and mathematics (85.7%), social sciences and humanities (88.2%) and health sciences (91.7%), but above the average in applied sciences and engineering (95.1%). Co-supervisors were also involved in 62.5% of cases when all disciplines were combined. Students in the sciences were also more likely to choose their supervisors before their research subject, while in the social sciences and the humanities the opposite is observed: students choose their research subject, and then find a supervisor who is interested to supervise a thesis on the topic.

Similar results for the domains of natural and social sciences were also observed by Delamont, Atkinson and Parry (1997) and Ridding (1996): supervisors were very active in the choice of research topics in the sciences. On the other hand, students in the social sciences were more independent in that respect, which is probably a consequence of the more important collaborative nature of the former group of disciplines. Different disciplines have indeed different manners of enrolling doctoral students into research teams, and doctoral students in the natural or medical sciences are more often part of research teams than their colleagues of the social sciences. As pointed out by Pole *et al.* (1997), "In team-based research the PhD student may have a role as a formal member of the team which means that he/she contributes to a sizeable research project which is likely to make a significant contribution to the discipline in which it is based" (p. 58). These links between the work of the supervisors and that of the students might, in turn, increase their students' participation in peer-reviewed papers.

Finally, these different 'supervisory' practices also have a strong influence on degree completion and on doctoral students' satisfaction of their graduate school experience. As shown by Smeby (2000), 73% of students in the natural sciences have completed their degrees by the end of their seventh year, while less than half of students in the humanities and social sciences had done the same. Again the participation of students into research teams was noted as a determining factor: while students in the 'hard' sciences were directly involved in their supervisors' research, those of the 'softer' social sciences and the humanities typically work alone. Walford (1981) comes to the same conclusion: "The proportion [of doctoral students dissatisfied with the supervision they receive] is generally found to be higher in the social sciences and arts than in the pure sciences, probably due to the much closer nature of supervision generally offered in the experimental sciences." (p. 147-148).

This participation in research groups benefits the student—as they have more opportunity to publish and to graduate—but also the advisors, as those who supervise graduate students on projects related to their own research were more productive than those who did not (Kyvik and Smeby, 1994). This effect was observed in the natural and medical sciences, but not in the social sciences and humanities. This can be explained by Kyvik and Smeby's (1994) and others' (Delamont, Atkinson and Parry; 1997; Pole *et al*., 1997; Ridding, 1996) observation that doctoral students' research was more often linked to that of their supervisors in the sciences than in the social sciences, or by the fact that co-authorship is less important in the latter group of disciplines than in the former (Larivière, Gingras and Archambault, 2006).

**Data and methods**

This paper uses the whole population of PhD students enrolled in Quebec's universities between 2000 and 2007 (N=27,393). Papers authored by these students during the 2000-2007 period were retrieved from Thomson Reuters' Web of Science (WoS) by matching the names of all of Quebec's doctoral students—obtained though an agreement with the *Ministère de l'éducation, du loisir et du sport du Quebec*—with the names of authors of papers with at least one institutional address from Quebec. Given that the WoS does not index the complete first name of the authors, but only their initials[1], this rough match generates a high number of false positives—papers authored by other researchers having the same name as a doctoral student (homographs)—which were removed using both manual and automatic validation.

---

[1] Although Thomson Reuters now index the full first name and initials of the authors and provide a link between each of the authors and its institution of affiliation, this was not available at the time the data for this thesis were compiled. Thomson Reuters does not yet provide this data retrospectively.

Automatic validation of students' publications was performed using the list of manually assigned publication of Quebec university professors (Gingras, Larivière, Macaluso and Robitaille; 2008 Larivière, Macaluso, Archambault and Gingras, 2010). Using the patterns found in the relationship between the discipline of professors' departments and the discipline of their publication, we created a simple algorithm which automatically assigned at least one paper to 80% of doctoral students. The remaining publications were manually assigned or rejected. To do that, we look closely and carefully at all papers contained in a file with a critical eye at their disciplines and their particular topics. The titles of the papers were often searched on the Internet to find the original paper where the complete name of the authors as well as the links between the authors and their institutional addresses could be found. After all of these steps, 31,738 author-article combinations were retained (out of the 313,367 originally obtained with the first match) for 25,159 distinct papers authored by 8,468 doctoral students. Numbers presented in this paper are a subset of this dataset, as they only take into account the papers published *during* doctoral studies.

Each of the ≈11,000 scientific journals indexed yearly in the WoS were categorized into four broad categories: health sciences (health), natural sciences and engineering (NSE), social sciences (SS), and arts and humanities (AH), which are used for the disciplinary breakdown of the numbers presented in this paper. Publication counts presented in this paper are based on the number of articles, notes and review articles to which authors from Québec contributed during the 2000-2007 period. Hence, editorials, book reviews, letters to the editor or meeting-abstracts are excluded from the analysis, as they are not generally considered as original contributions to scholarly knowledge (Moed, 1996). These numbers are based on full counting of papers, as opposed to fractional counting sometimes used in bibliometrics. Hence, each individual or organization contributing to a paper is assigned one 'full' contribution, instead of a fraction of a contribution, irrespective of their rank in the author order. Papers are considered as being authored by doctoral students when at least one of the authors is enrolled in a PhD program in one of Quebec's universities during the publication year of the paper or has been enrolled during the year prior to the publication of the paper. In other words, in line with Lee (2000), doctoral students' papers are still considered as such until one year after they graduate or leave the program. Note that this practice is also found in the large teams of particle physics, where researchers get to sign the papers coming out of the experiment until one year after they leave the team (Biagioli, 2003). The distribution of PhD students' papers as well as the distribution of PhD students in each of the four disciplines is presented in Figure 1. It shows that SS is the area with the higher proportion of students, followed by NSE, health and then AH. In terms of number of papers, health obtains the greatest proportion, followed by NSE, SS and then AH.

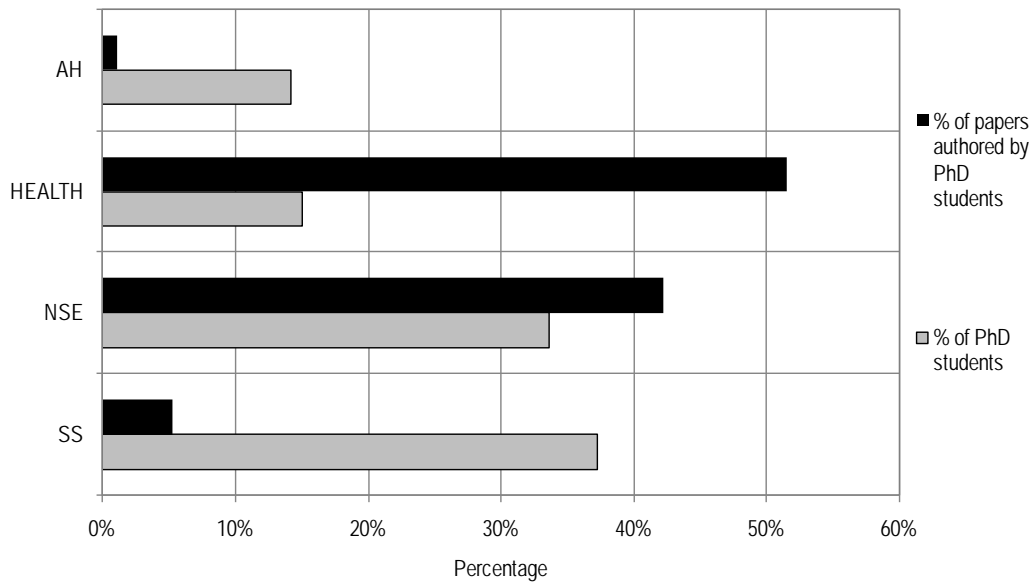

Figure 1. Percentage of Quebec's university papers to which doctoral students' contributed and percentage of doctoral students and professors who published at least one paper, 2000-2007

Citations measures are made over the full period, which means that the total number of citations received since publication by a given paper are counted. In order to compare data between different specialties, each article's number of citation is divided by the average number of citations received by papers of the same discipline published the same year (Moed, De Bruin and van Leeuwen, 1995; Braun and Schubert, 1986)[2]. Self-citations are excluded and impact measures are normalized so thats when the average of relative citations (ARC) is above 1, the articles of the group of researchers in a given field are, on average, cited above the world average for the same field. Conversely, an ARC below 1 means that the number of citations received is below the world average. D

---

[2] Document types—research article or review—have not been taken into account in the normalization process.

Finally, the well known limitations of bibliometrics apply to this analysis. Indeed, the measures presented here do not include all documents likely to have been published by doctoral students, as no bibliometric database indexes all of the scholarly literature published worldwide. This limitation is more important in the social sciences and the humanities, where the application of article counting methods poses two main problems: 1) no coverage of research output in media other than journal articles, 2) very limited coverage of research output in the form of articles written in languages other than English (Archambault *et al.,* 2006). The first limitation is attributable to discipline-specific dissemination media—scholars tend to publish fewer articles and more books in the social sciences and humanities (SSH) (Larivière *et al.*, 2006) or more conference proceedings in computer sciences (Lisée, Larivière and Archambault, 2008)—while the second is attributable to SSH scholars' preferred language of knowledge dissemination and their research interests (Gingras, 1984). The extent of these limitations varies considerably by discipline. Some social sciences, such as economics, administration and psychology are less local and are more international and thus have a core of international journals and the scholarly article plays a more central role in these disciplines. In other words, this research method does not make it possible to gather data on each publication a graduate student might have signed. However, it focuses on the contributions that were made in core journals of their respective fields (Garfield, 1990), that is, journals that are most cited in other journals.

One indicator used in this paper could be influenced by these differences in coverage: the percentage of PhD students with at least one paper. Indeed, the lower proportion of PhD students with papers in disciplines of the social sciences and humanities we can anticipate might be due to the a lower coverage by the WoS of their output, compared with students of the natural and medical sciences. In order to take into account these differences in coverage, we compare PhD students' results with those obtained by the entire population of professors of Quebec universities (Larivière, Macaluso, Archambault and Gingras, 2010) of the same disciplines, which are as influenced by coverage issues as students are. This provides a baseline against which we can compare students' output.

**Results and discussion**

*PhD students' contribution to the advancement of knowledge*

Figure 2 presents the percentage of Quebec's university papers to which at least one PhD student contributed as well as the percentage of PhD students and professors who published at least one paper during the 2000-2007 period. It shows that 63% of doctoral students in health and 40% of those in NSE have contributed to at least one paper during their doctorate. On the other hand, about 10% and 4% of students in SS and AH, respectively, have done so. Although one could argue that these differences are artefacts caused by the coverage of the WoS, the comparison of PhD students' results with those obtained by faculty members shows that students in the AH and SS are, respectively, 7.4 and 3.8 times less likely to author a paper than faculty members, while these percentages are of 2.0 and 1.1 in NSE and health, respectively. This provide clear evidence, independent from coverage issues—by which faculty members are as influenced as PhD students—, that PhD students in the AH and SS are less likely to be involved into publication-related research activities during their doctorate than their colleagues of the NSE and health. Similarly, these findings are consisted with those provided by surveys—which, by definition, 'index' the entire production of respondents (Nettles and Millett, 2006; Gemme and Gingras, 2008)[3].

The tendency is similar in terms of their overall proportion of the output of the province. Indeed, for both health and NSE, about 30% of all Quebec university papers have doctoral students as authors or co-authors. On the other hand, a smaller proportion of the province's papers are authored by doctoral students in SS (19%) and AH (13%). Several factors can explain these disciplinary differences. First and foremost, participation in peer-reviewed papers is increasingly considered, in the disciplines of the medical and natural sciences, as one of the requirements for the completion of the doctoral degree—since theses in these disciplines often taken the form of a series of articles. Even though data on the extent of this practice is scarce[4], it seems that the 'standard' monograph thesis has been replaced, in most of these scientific disciplines, by a series of articles which have to be published in peer-reviewed journals (Breimer, 2010; Holdaway, 1994).. Hence, it is thus normal that in these disciplines, doctoral students publish scientific papers, since it is now often considered to be one of the requirements for degree completion because of the form of the thesis. On the other hand, in the disciplines of the social sciences and of the humanities, the format of the thesis is still that of the monograph—which is similar to the book format often used by researchers for disseminating research. As a consequence, the writing of articles is often considered less important in their research training.

---

[3] Let us recall here that the study of Gemme and Gingras (2008), also based Quebec, has shown that only 2% of students have authored a book, and 5% to a book chapter.
[4] A report for the UK Council for Graduate Education shows that the prevalence 'PhDs by published works' has increased by more than 100% between 1996 and 2004 (Powell, 2004).

Another aspect is the structure of the student-supervisor relationship. In the social sciences and the humanities, students typically work from home, often with less interaction with their supervisors than in the natural sciences and medicine (Delamont, Atkinson and Parry, 1997; Pole *et al.*, 1997; Ridding, 1996). They also tend to be only remotely involved in their advisors' research, as most professors do not have any team of their own and prefer to perform research alone (Larivière, 2007). In that sense, it is normal that the highest proportion, among the SSH disciplines, of doctoral students contributing to papers is observed in the discipline of social sciences, as the use of quantitative methods allows more opportunities for collaboration (Moody, 2004). The positive effect of collaboration was also observed by Louis *et al.* (2007).

On the other hand, in the natural and medical sciences, students typically go to the lab every day, and get to work, in addition to their own doctoral research, on other researchers' projects (other students, post-doctoral fellows, professors, etc.). Moreover, their research is often one component of a bigger project on which other researchers in the group work, and students typically have difficulty distinguishing their own work from that of their supervisors (Pole *et al.*, 1997).

Hence, if one uses the participation in peer-reviewed papers as an indicator, doctoral students in the natural and medical disciplines are more socialized and integrated to research—both their own as well as that of other researchers in the lab—than their colleagues of the social sciences and humanities.

Along the same lines, Pontille (2004) has showed that the contribution needed to sign a scientific paper varies greatly by discipline. In the habitual team structure of the natural and medical sciences, senior co-authors habitually grant authorship to junior staff involved in various parts of the research. On the other hand, in disciplines of the social sciences and humanities, because of the historical authorial figure, established researchers are less disposed to grant co-authorship to subordinates—even when they participated to the research (Pontille, 2004). These different authorship attribution methods might explain, at least in part, the differences observed.

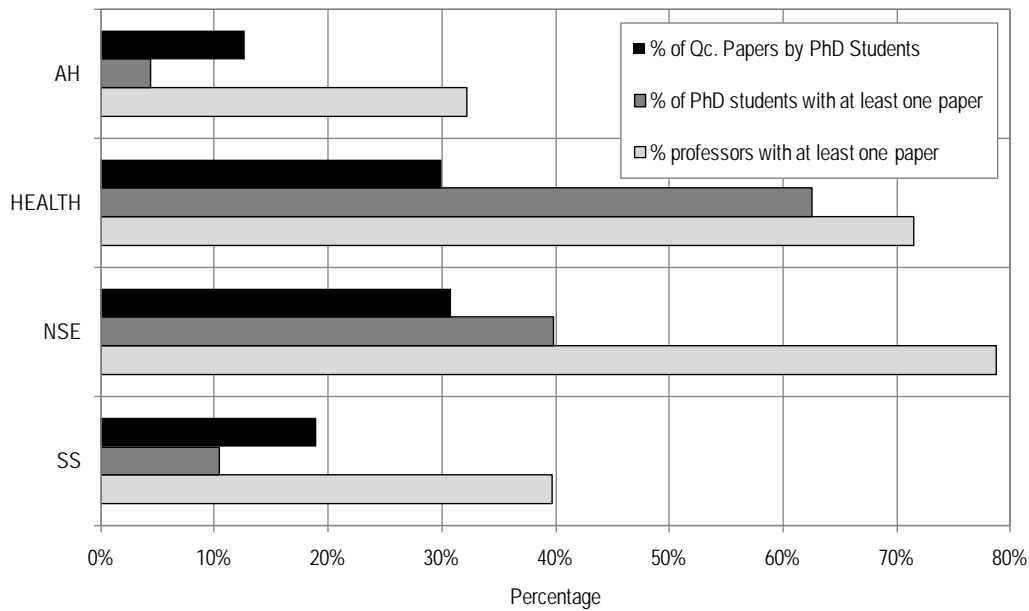

Figure 2. Percentage of Quebec's university papers to which doctoral students' contributed and percentage of doctoral students and professors who published at least one paper, 2000-2007

*PhD students' collaboration patterns*

Many studies have shown that, in most disciplines, the quasi-totality of the papers is co-authored (see, among others, Cronin, 2005; Larivière, Gingras and Archambault, 2006). Doctoral students' proportion of co-authored papers is even higher, with about 99% of papers having more than one author in NSE and health (not shown). As one could expect, only the humanities display a different pattern: only 10% of the papers are co-authored, with only a very small difference between doctoral students' papers and other Quebec papers. Along these lines, Figure 3 provides data on the average number of authors of doctoral students' papers and of other Quebec papers to which no doctoral student contributed. It reveals that the mean number of authors is higher for papers to which doctoral students have contributed than for other Quebec papers, as if the PhD student was an 'extra' author added. More specifically, for all disciplines combined, papers to which doctoral students contributed have twice as many authors than Quebec papers to which they did not contributed (18.9 vs. 9.8). This important difference is mainly due to NSE—and particularly physics—where there are, on average, more than 34 authors on PhD students' papers against 23 on Quebec papers to which they did not contributed. We also observe a similar pattern the three other areas, although the difference between the two groups of paper is smaller. This strongly suggests that, in their apprenticeship, doctoral students benefit from the help of other researchers who are more experienced and established. Even if we do not have any information on the link between the student and the other authors, it could be expected, given the qualitative research on doctoral students' socialisation to research, that these co-authors are, more often than not, their supervisors/mentors (Pole *et al.*, 1997).

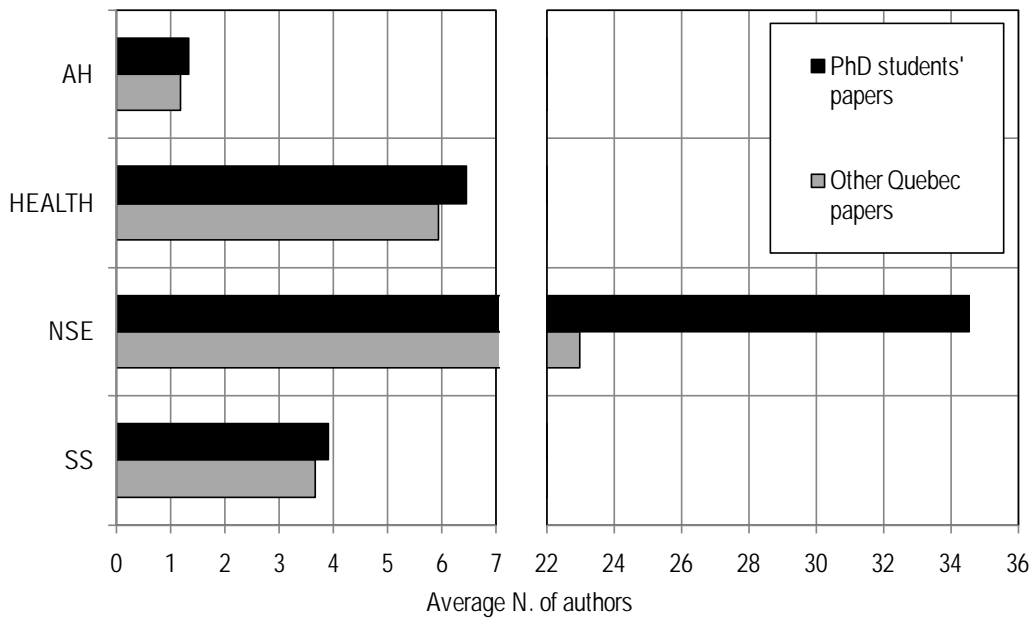

Figure 3. Average number of authors of papers to which doctoral students contributed and of Quebec's other papers, by discipline, 2000-2007

Science is a collective endeavour, and collaborators are increasingly found in foreign countries. Indeed, researchers throughout the world are increasingly collaborating with foreign partners (see, among others, Georghiou, 1998; Glänzel, 2001; Leydesdorff and Wagner, 2008) and Quebec researchers are no exception (Larivière, 2007). All collaboration measures presented so far in this section showed that doctoral students' papers were more likely to be the result of collaboration. Figure 4 provides a distinct picture and presents the percentage of doctoral students' papers and of other Quebec papers that are authored in collaboration with foreign colleagues, by discipline of the journal in which the papers are published. Indeed, in all disciplines but the arts and the humanities, a significantly lower proportion of doctoral students' papers than of other Quebec papers have international co-authors. Hence, even if doctoral students' integration to research is generally made in bigger teams, these teams are less prone to include international colleagues. One could think that, given their apprenticeship status in research, doctoral students might not be attributed international projects by their supervisors in natural sciences and medicine, or don't have yet international contacts to publish with in the SSH.

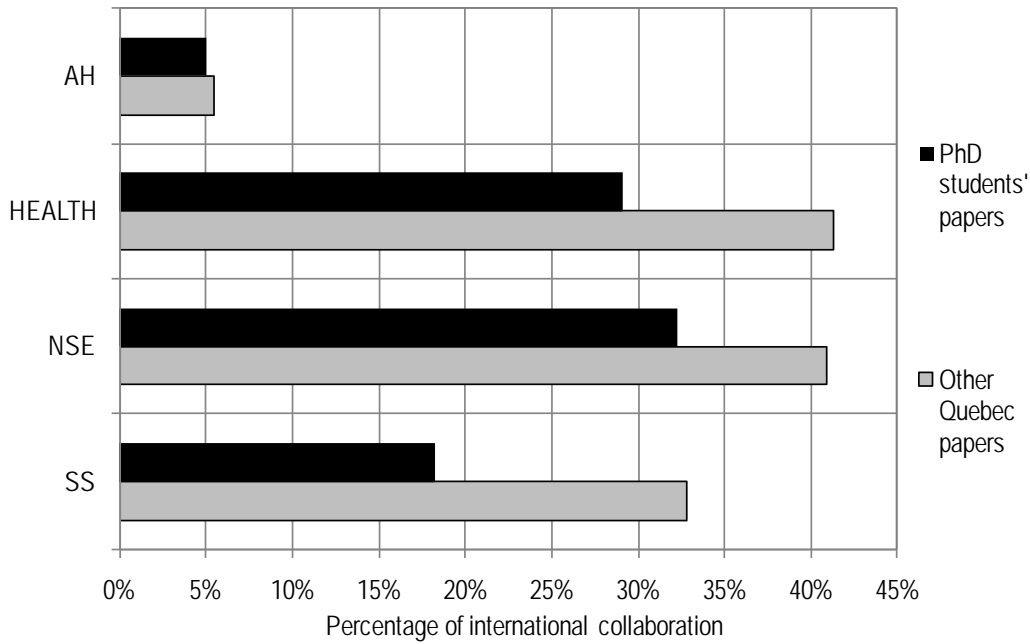

Figure 4. Percentage of international collaboration of papers to which doctoral students contributed and of Quebec's other papers, by discipline, 2000-2007

*Degree completion and post-graduation research output*

As shown in Figure 2, doctoral students do not contribute equally to the output. Indeed, a very low proportion of PhD students in the AH (4%) and in SS (10%) have been authors or co-authors of papers during their doctorate, while 40% of students in NSE and 63% of those in health have published at least one paper during their doctorate. Publishing is an indisputable effect of socialization to research and several authors have analyzed the impact of socialization to research on students' ulterior research careers. More specifically, authors such as Turner and Thompson (1993) and Gardner (2007) have positively linked socialization to degree-completion. Nettles and Millets (2006) established that research productivity was positively linked with degree completion and time spent in the program, i.e., students with a publication record did not spend more time in the program. Similarly, Seagram, Gould and Pyke (1998) observed that collaborating with doctoral advisees on conference papers was one of the factors reducing time to completion.

Figure 5 presents the number of papers by doctoral students of the 2000, 2001 and 2002 cohorts, for those who completed their doctorate as well as those who had not completed it as of the end of 2007 (N=6,596). It clearly demonstrates that, in each of the disciplines, those who had completed their doctorate published a higher number of papers than those who had not completed the program. These data provide strong evidence of the links between publication activity and degree completion. Indeed, an important aspect of the doctorate is to contribute to the advancement of scholarly knowledge in a discipline. It is thus normal that, by publishing papers—which are contributions to knowledge—doctoral students increase their chances of completing their doctoral degrees. On the other hand, the average impact per paper is not significantly different for the two groups of students (not presented here), if one excludes doctoral students who have not published any paper.

Although one could argue that, in the natural sciences and medicine, these differences are caused by theses which consist of a series of published articles—which is now the standard form of the thesis in these disciplines—these practices seldom exist in the social sciences and humanities and, hence, cannot explain the differences observed in these disciplines, which are as high as those observed in the natural sciences and medicine. On the whole, being integrated into research is strongly related with doctoral students' degree completion.

Figure 5. Average number of papers by doctoral students having completed their program and by doctoral students that have not completed their program as of the end of 2007, for 2000-2002 cohorts (N=6,596)

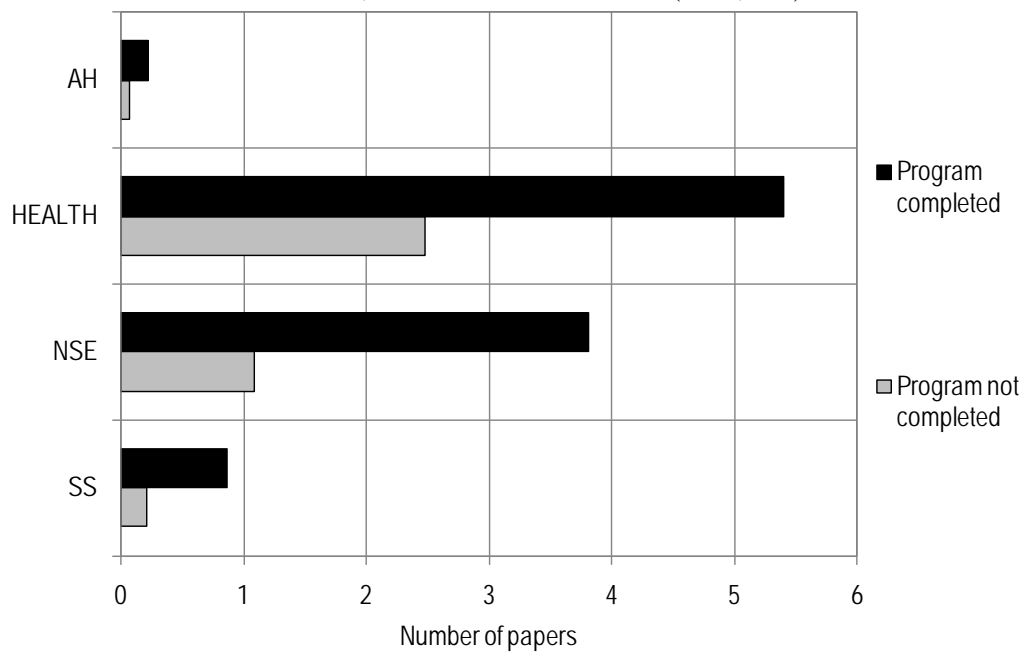

To measure the relationship between pre-graduation integration into research and post-graduation research activity, we compiled the average number of post-graduation papers as a function of classes of pre-graduation papers (Figure 6). In order to have enough numbers of pre- and post-graduation papers, only the subset of doctoral students who had graduated in 2003 or 2004 are considered (N=2,319 PhD students).

Figure 6 clearly shows that there is a positive relationship between the two variables; those who publish more during their doctorate are more likely to publish more afterwards[5]. We observe that in all disciplines but social sciences and humanities—where the trends are less clear—doctoral students who had not authored any paper during their studies obtain lower publication rates after graduation. This is an obvious effect of socialization and integration into research: students who are more involved in research during their doctorate are socialized to the publication *habitus* (Bourdieu, 1980; 2001) and keep this *habitus* after graduation, when they themselves become members of the scientific community. This also suggests that those who have been involved in research during their doctorate have higher probabilities of obtaining research positions (postdoctoral fellows, researchers or professors) after graduation.

---

[5] One of the limitations of this figure is that it only includes post-graduation papers having at least one Quebec address, as this was one of the matching criteria for assigning the student's papers. Hence, papers authored during a post-doctoral fellowship abroad are not included, except when they are written with collaborators from Quebec. This thus reduces the probability of finding a post-graduation measurable output for any of the classes of pre-graduation productivity.

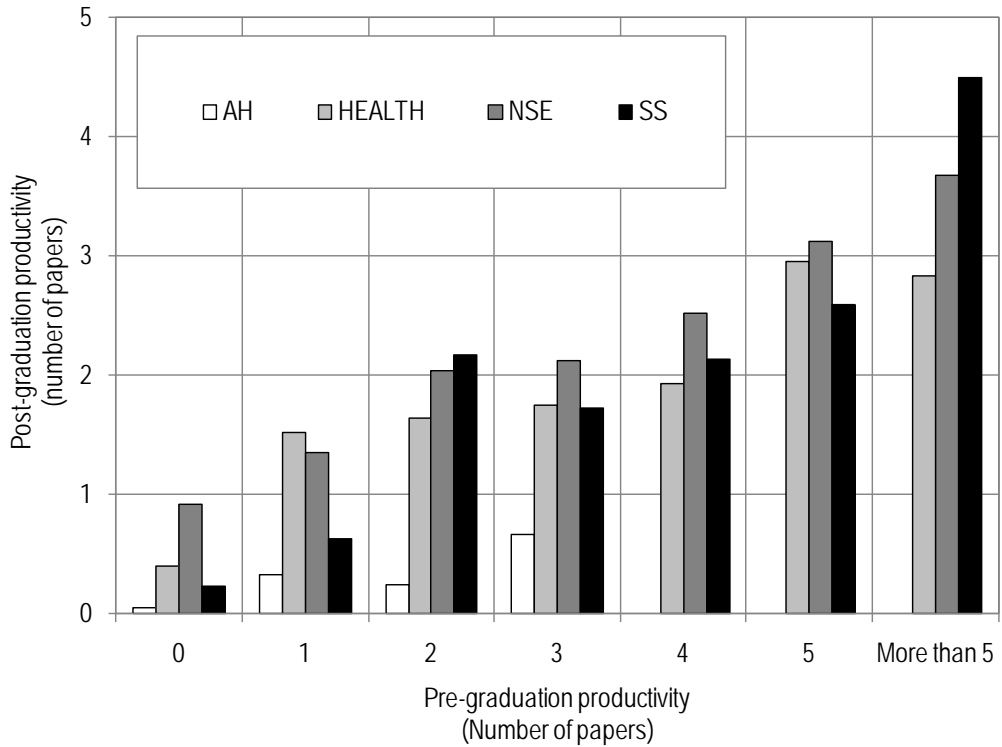

Figure 6. Relationship between pre-graduation productivity and post-graduation productivity, for the subset of doctoral students who graduated in 2003-2004 (N=2,319)

*PhD students' scientific impact*

Several studies have analyzed the relationship between the aging of researchers' and their scientific impact or creativity (see the recent review by Feist, 2006). Although these studies offer diverging results, two general trends emerge from the data: older researchers publish more papers (as they are on top of the hierarchy) but younger ones typically have a higher scientific impact (as they are, it is assumed, at the top of their creativity). None of the existing studies, however, offer any data on the scientific impart of doctoral students who are, in addition to generally being of a young age, are also, in a sense, prior to the year '0' of their careers—as the year of completion of the PhD degree is generally considered as the start of an academic career. One might thus wonder—since we know that younger professors' scientific impact is higher in Quebec (Gingras, Larivière, Macaluso and Robitaille, 2008)—if doctoral students' papers are having higher scientific impact than other Quebec papers.

Figure 7 presents scientific impact measures (ARC) of papers to which doctoral students contributed, as well as for all other Quebec papers. In health, SS and AH, PhD students' papers obtain lower citation rates. On the other hand, papers in NSE with doctoral students as co-authors obtain significantly higher citation rates. There is, thus, an important disciplinary component to this scientific impact or "creativity" of researchers, as students' papers in more empirical disciplines of NSE obtain a higher number of citation while those in more theoretical disciplines of SS and AH obtain lower citation counts. Theoretical creativity might take more maturity and a superior knowledge of previous literature, which only comes with age while empirical creativity might be more a function of technical skills and curiosity, two characteristics generally found in doctoral students.

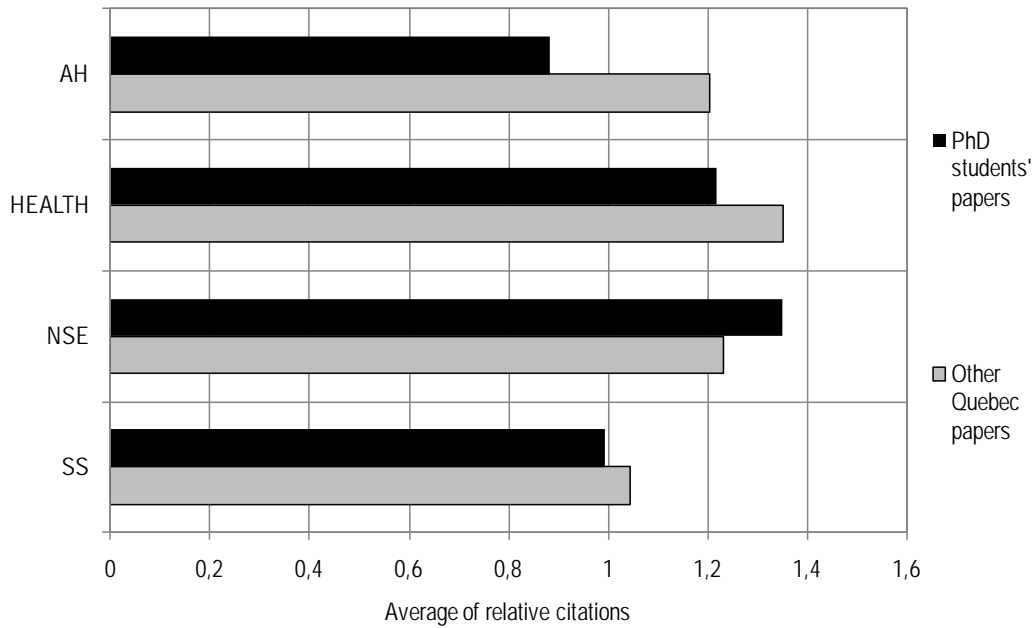

Figure 7. Average of relative citations of papers of papers to which doctoral students contributed and of Quebec's other papers, by discipline, 2000-2007

**Conclusion**

This paper provided the first large scale analysis of doctoral students' publication activity. It highlighted the essential role of doctoral programs not only for the 'reproduction' of researchers, but also for the research system, as doctoral students contribute, during their studies, to a considerable proportion of the new knowledge being created, especially in health and NSE disciplines. It also revealed that the percentage of doctoral students who are involved in research—as measured by their involvement in publications—varies a great deal among disciplines, with a high of 63% in health, but a low of 5% in AH. Even if the disciplinary classification used in this paper is not exactly the same as that of Nettles and Millets (2006)—the most comprehensive source on doctoral students' publications to this date—we see that the 'spectrum' of doctoral students' participation in papers by discipline is quite similar, with students in the sciences publishing more papers and students in education publishing less papers. A similar trend is observed in Lee's (2000) data. Several factors explain these differences, among which the different formats of doctoral theses (article-based vs. monograph) as well as the various modes of organization of research are probably the most important. Despite the fact that we do not have data on the extent of article-based theses and of its differences among disciplines, we know that this 'form' of thesis has become quite frequent in the natural and medical sciences (Breimer, 2010; Holdaway, 1994) and, hence, have the obvious effect of increasing the contribution of doctoral students to papers in these disciplines.

Along these lines, in the natural and medical disciplines collaboration is the preferred mode of production of knowledge (Larivière, Gingras and Archambault, 2006) and doctoral students' research projects are de facto linked with those of their supervisors (Gemme and Gingras, 2008). Doctoral students, then, become part of the research team and work, along with the other members of the team, on the various projects of the lab, of which their own doctoral research is only one component (Pole et al., 1997). This integration in a research team takes away the burden of socialization and integration on the sole shoulders of the advisors, as several other actors (postdoctoral fellows, laboratory technicians, etc.) are involved in the process (Delamont, Atkinson and Parry, 1997; Gemme and Gingras, 2008), and increases the opportunity for doctoral students to participate in various research projects. However, in the social sciences and humanities—where research teams are not the norm (Larivière. Gingras and Archambault, 2006)—doctoral students are less integrated into their advisors' research (Delamont, Atkinson and Parry, 1997; Ridding, 1996), as professors are either less likely to need help in their research or the—often technical—work the students perform does not grant authorship. They typically work as assistants on their advisors projects—which may or may not be linked with their thesis—for a few days per week and then work in their own projects the rest of their time (Legault, 1993).

Even though we cannot assess the direction of the relationship, publishing papers during the doctorate is positively linked with PhD students' degree completion and with post-graduation research productivity. In all disciplines—even those of the SS and AH where only a small proportion of doctoral students publish—students involved in the publication of scientific papers are, by far, more likely to complete their doctorate faster. This fact is not without having policy implications on the training of tomorrow's researchers. Doctoral programs, aiming at the training of new researchers, should focus on the optimal integration of students during the course of the program. In that respect, doctoral training in the NSE and health is quite efficient, as doctoral students collaborate with their supervisors and are already doing what they will do, if they obtain a research position after they graduate, that is: publish papers. This is exemplified by the fact that most doctoral theses in these disciplines have the form of a series of articles. In the SS and AH on the other hand, professors do not generally have research teams and are, thus, less prone to involve PhD students in their own research projects. Doctoral students in SS and AH are, thus, generally left on their own, and it is undisputedly more difficult to keep an original research project going on if one is alone rather than part of a team. Although we do not propose that PhD students' formation in SS and AH mimics that of NSE and health—as both the methods used and the objects studied are not always appropriate for collaboration—this paper provides clear evidence that this lesser inclusion into collective research affects time to completion: while more than 50% of NSE and health students of the 2000-2002 cohorts had completed their program as of 2007, this percentage is half smaller (25%) in SS and AH. Though no panacea can solve the problem of time to completion, it remains that a better integration of doctoral students into the collective dynamics of research would yield better individual and collective results.

**Acknowledgements**

The author wishes to thank professors Yves Gingras, Jamshid Beheshti, France Bouthillier, John Leide and Pierre Pluye for their careful reading of the doctoral thesis from which this paper is drawn, as well as the four anonymous referees for useful comments and suggestions. Special thanks to Benoit Macaluso, Mario Rouette, Philippe Mirabel and Benoit Gagné for their help in the assignation of doctoral students' papers. Finally, support from the Social Sciences and Humanities Research Council of Canada through the CGS Doctoral Scholarship program is also acknowledged.